\documentclass[aps,noshowpacs,nobalancelastpage,floatfix,onecolumn,final]{revtex4}
\usepackage{amssymb}
\usepackage[dvips]{graphicx}

\begin{document}

\title{Measurement-Induced Entanglement for Excitation Stored in Remote
Atomic Ensembles}
\author{C. W. Chou, H. de Riedmatten, D. Felinto, S. V. Polyakov, S. J. van
Enk,$^{\dagger }$ and H. J. Kimble}
\affiliation{Norman Bridge Laboratory of Physics 12-33\\
California Institute of Technology, Pasadena, CA 91125\\
$^{\dagger }$Bell Labs, Lucent Technologies, Room 1D-428\\
600-700 Mountain Ave, Murray Hill, NJ 07974}
\date{\today}
\abstract{A critical requirement for diverse applications in Quantum
Information Science is the capability to disseminate quantum
resources over complex quantum networks \cite{QC,arda}. For example,
the coherent distribution of entangled quantum states together with
quantum memory to store these states can enable scalable
architectures for quantum computation \cite{chuang03}, communication
\cite{briegel98}, and metrology \cite{lloyd05}. As a significant
step toward such possibilities, here we report observations of
entanglement between two atomic ensembles located in distinct
apparatuses on different tables. Quantum interference in the
detection of a photon emitted by one of the samples projects the
otherwise independent ensembles into an entangled state with one
joint excitation stored remotely in }$10^{5}$ atoms at each site
\cite{duan01}. After a programmable delay, we confirm entanglement
by mapping the state of the atoms to optical fields and by measuring
mutual coherences and photon statistics for these fields. We thereby
determine a quantitative lower bound for the entanglement of the
joint state of the ensembles. Our observations provide a new
capability for the distribution and storage of entangled quantum
states, including for scalable quantum communication networks
\cite{duan01}.} \maketitle

Entanglement is a uniquely quantum mechanical property of the
correlations among various components of a physical system. Initial
demonstrations of entanglement were made for photon pairs from the
fluorescence in atomic cascades \cite{clauser78,aspect82} and from
parametric down conversion \cite{mandel-wolf95}. More recently,
entanglement has been recognized as a critical resource for
accomplishing tasks that are otherwise impossible in the classical
domain \cite{QC}. Spectacular advances have been made in the
generation of quantum entanglement for diverse physical systems
\cite{QC,arda}, including entanglement stored for many seconds in
trapped ions for distances on the millimeter scale
\cite{wineland,blatt}\textit{,} long-lived entanglement of
macroscopic quantum spins persisting for milliseconds on the
centimeter scale \cite{julsgaard01}, and remote entanglement carried
by photon pairs over distances of tens of kilometers of optical
fibers~\cite{entangle-km}.

For applications in quantum information science, entanglement can be
created deterministically by way of precise control of quantum
dynamics for a physical system, or probabilistically by way of
quantum interference in a suitable measurement with random instances
of success. In the latter case, it is essential that success be
heralded unambiguously so that the resulting entangled state is
available for subsequent utilization. In either case, quantum memory
is required to store the entangled states until they are required
for the protocol at hand.

There are by now several examples of entanglement generated
\textquotedblleft on demand,\textquotedblright\ \cite{QC}
beginning with
the realization of the EPR paradox for continuous quantum variables \cite{ou92}
and the deterministic entanglement of the discrete internal
states of two trapped ions \cite{turchette98}. Important progress
has been made towards measurement-induced entanglement on various
fronts, including the
observation of entanglement between a trapped ion and a photon \cite{blinov04}.

Against this backdrop, here we report the initial observation of
entanglement created probabilistically from quantum interference in
the measurement process, with the resulting entangled state heralded
unambiguously and stored in quantum memory for subsequent
exploitation. As illustrated in Fig. 1, the detection of a photon
from either of two atomic ensembles (\textit{L}, \textit{R}) in an
indistinguishable fashion results in an entangled state with one
\textquotedblleft spin\textquotedblright\ excitation shared at a
distance of $2.8$\ m between the ensembles and distributed
symmetrically among $\sim 10^{5}$\ atoms at each site \cite{duan01}.
Confirmation of entanglement is achieved by mapping this stored
excitation onto light fields after a $1$\ $\mu $s delay
\cite{duan01,felinto05} and by suitable measurements of the quantum
statistics of the resulting optical fields. Our results provide the
first realization of the capability to transfer a stored entangled
state of matter to an entangled state of light.

Our experiment is motivated by the
protocol of Duan, Lukin, Cirac and Zoller (DLCZ) \cite{duan01} for
the realization of scalable quantum communication networks with
atomic ensembles. The DLCZ protocol introduced a number of novel
ideas for quantum information processing and is the subject of
active investigation. In this direction, nonclassical
correlations~\cite{felinto05,kuzmich03,vanderwal03,jiang04,chou04,eisaman04,polyakov04,balic05}
and entanglement~\cite{kuzmich05} have been observed between pairs
of photons emitted by a single atomic ensemble. Observations of
coherence between two cylindrical volumes of cold Rubidium atoms
within a single magneto-optical trap have also been
reported~\cite{matsukevich04}, although entanglement was not
demonstrated between the two regions~\cite{vanenk05,matsukevich05}.

A simple schematic of our experiment is given in Fig. 1, with
further details provided in Refs.
\cite{felinto05,chou04,polyakov04}. For the writing stage of the
protocol, two classical pulses traverse the \textit{L}, \textit{R}
ensembles in parallel and generate fields $1_{L},1_{R}$ by
spontaneous Raman scattering (see Fig. 1a). The
intensity of the pulses is made sufficiently weak such that the
probability of creating more than one excitation in the symmetric
collective  mode \cite{duan01} of the ensemble is very low
\cite{chou04}.

\vspace{-1cm}
\begin{figure}[ht]
\begin{center}
\includegraphics[width=9cm,angle=270]{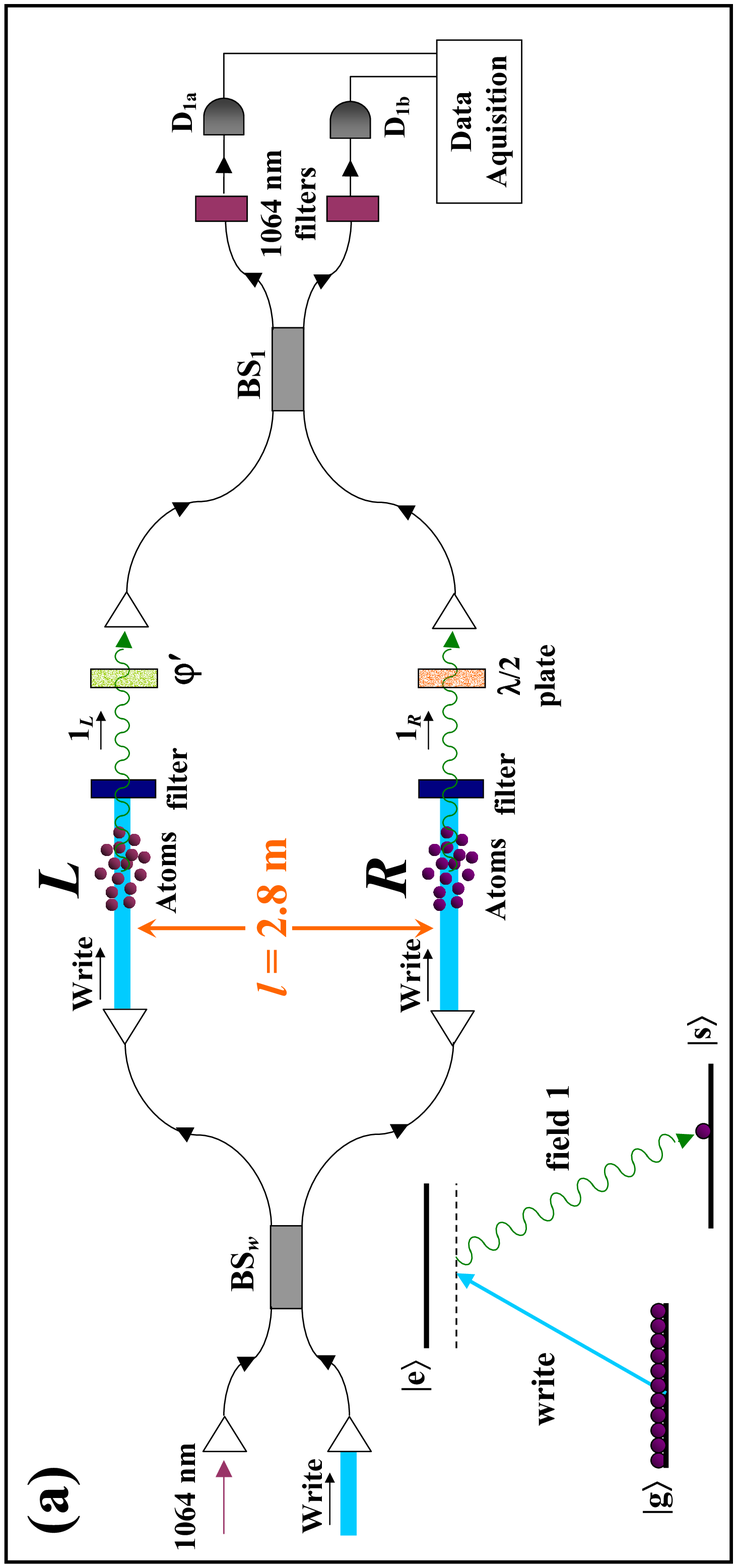}\\[0pt]
\vspace{-2.62 cm} \hspace{0.13cm}\includegraphics[width=9cm,angle=270]{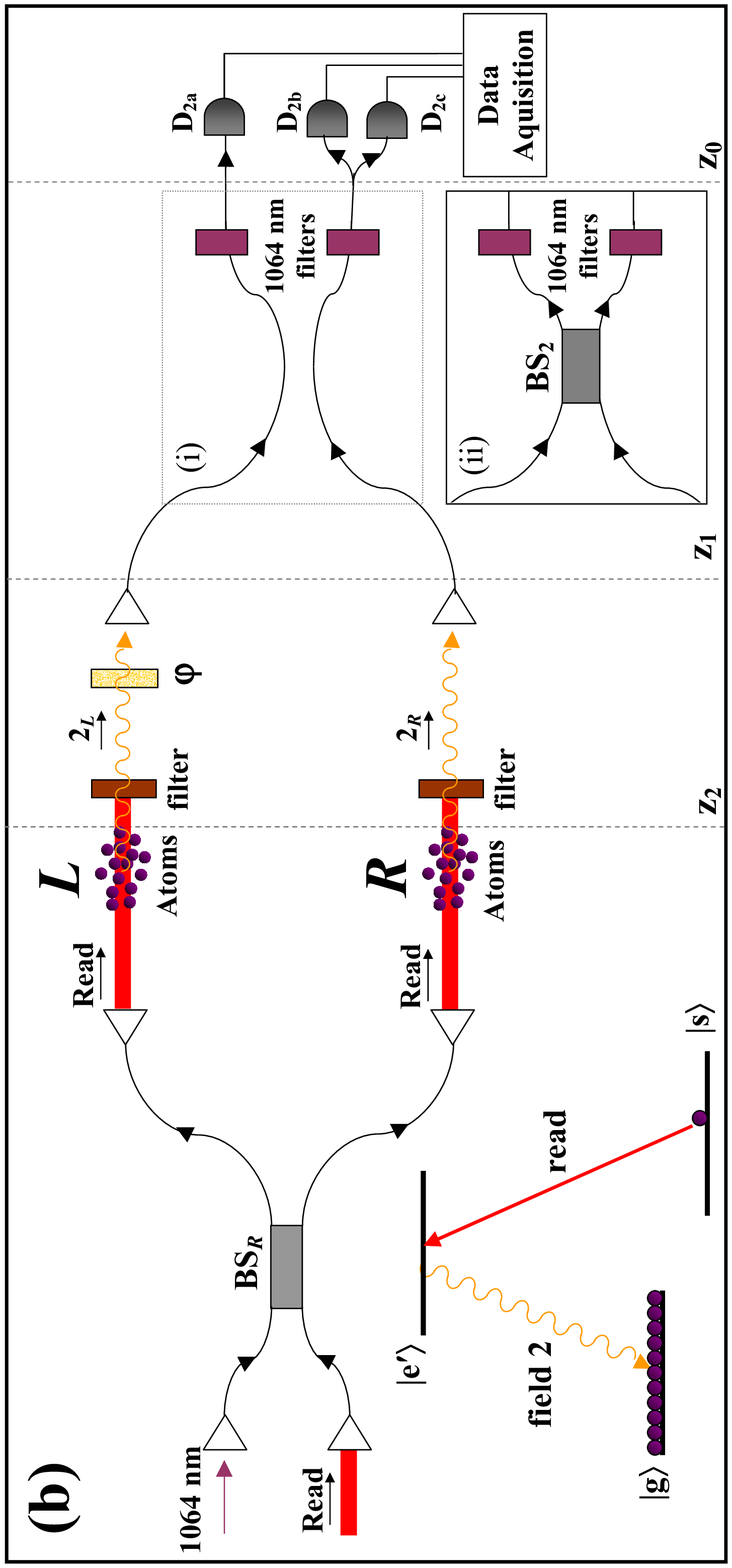} \\[0pt]
\end{center}
\vspace{-2.0cm}
\caption{An overview of our experiment to entangle two atomic
ensembles is shown. (a) Setup for generating entanglement between
two pencil-shaped ensembles $L$ and $R$ located within spherical
clouds of cold Cs atoms. The atomic level structure for the writing
process consists in the initial ground state $|g\rangle
$($6S_{1/2},F=4$ level of atomic Cesium),  the ground state
$|s\rangle $ for storing a collective spin flip ($6S_{1/2},F=3$
level), and  the excited level $|e\rangle $ ($6P_{3/2},F^{\prime
}=4$). The transition $|g\rangle \rightarrow |e\rangle $ in each
ensemble is initially coupled by a write pulse detuned from
resonance to generate the forward-scattered anti-Stokes field $1$
from the transition $|e\rangle \rightarrow |s\rangle $. The $L,R$
ensembles are excited by synchronized writing pulses obtained from
beam splitter BS$_{w}$. After filtering, the Stokes fields
$1_{L},1_{R}$ are collected, coupled to fiber-optic channels, and
interfere at beam splitter $BS_{1}$, with outputs directed towards
two single-photon detectors $D_{1a},D_{1b}$. (b) Schematic for
verification of entanglement between the \textit{L}, \textit{R}
ensembles by conversion of atomic to field excitation by way of
simultaneous read pulses obtained from  BS$_{R}$. The read pulses
reach the samples after a programmable delay from the write pulses,
and couple the transition $|s\rangle \rightarrow |e^{\prime}\rangle
$ ($|e^{\prime}\rangle$ being the $6P_{1/2},F^{\prime }=4$ level),
leading to the emission of the forward-scattered Stokes fields
$2_L,2_R$ from the transition $|e^{\prime }\rangle \rightarrow
|g\rangle $. The inset (i) shows the configuration used to measure
the diagonal elements $p_{ij}$ of
$\tilde{\protect\rho}_{2_{L},2_{R}}$ in Eq. \protect\ref{rhotilde}
from the photo-detection events at $D_{2a},D_{2b},D_{2c}$.
Reconfiguring the fiber connections we can easily pass from the
configuration of inset (i) to the one of inset (ii), which is used
to generate interference of the $2_{L},2_{R}$ fields at beam
splitter $BS_{2}$ to measure the off-diagonal coherence $d$ in
$\tilde{\protect\rho}_{2_{L},2_{R}}$. In (a) and (b), the incident
write and read beams are orthogonally polarized and combined at a
polarizing beam splitter, and are focused to a waist of about $30\mu
m$ in the sample region. All beam splitters BS are polarization
maintaining fiber beam splitters. The $\approx$ 12m arms of both
write and read interferometers are actively stabilized using an
auxiliary Nd:YAG laser at 1.06~$\mu$m.} \label{fig1}
\end{figure}

Entanglement between the \textit{L}, \textit{R} ensembles is created by
combining the output fields $1_{L},1_{R}$ on the beam splitter
$BS_{1}$, with outputs directed to two photodetectors
$D_{1a},D_{1b}$ (see Fig.1a). For small excitation
probability and with unit overlap of the fields at $BS_{1}$, a
detection event at $D_{1a}$ or $D_{1b}$ arises
indistinguishably from either field $1_{L}$ or $1_{R}$, so that the \textit{L%
}, \textit{R} ensembles are projected into an entangled state, which
in the
ideal case can be written as \cite{duan01,duan02}
\begin{equation}
|\Psi _{L,R}\rangle =\epsilon _{L}|0\rangle _{L}|1\rangle _{R}\pm
e^{i\eta _{1}}\epsilon _{R}|1\rangle _{L}|0\rangle _{R}\,,
\label{LRent}
\end{equation}
where $|0\rangle _{L,R}, |1\rangle _{L,R}$ refers to the two
ensembles $L,R$ with 0,1 collective excitations respectively, $\epsilon
_{L}$($\epsilon _{R}$) is the normalized amplitude of photon
generation from ensemble \textit{L}(\textit{R}), and the sign $\pm $
is set by whichever detector records the event. The phase $\eta
_{1}=\Delta \beta _{w}+\Delta \gamma _{1}$, where $\Delta \beta
_{w}$ is the phase difference of the write beams at the
\textit{L},\textit{\ R} ensembles, and $\Delta \gamma _{1}$ is the
phase difference acquired by the $1_{L},1_{R}$\ fields in
propagation from the ensembles to the beam splitter $BS_{1}$. Note that to achieve entanglement as in Eq. \ref{LRent}, $\eta_1$ has to be kept constant from trial to trial.

To verify the entanglement, we map the delocalized atomic excitation
into a field state by applying simultaneously strong read beams at
the two ensembles (see Fig.1b). If the state transfer were to
succeed with unit probability, the conditional state $|\Psi
_{L,R}\rangle $\ of the ensembles would be mapped to an entangled
state of two modes for the Stokes fields $2_{L},2_{R}$\ given in the
ideal case by \cite{duan01,duan02}
\begin{equation}
|\Phi _{LR}\rangle = \epsilon _{L}|1\rangle _{2_{L}}|0\rangle
_{2_{R}}\pm e^{i(\eta _{1}+\eta _{2})}\epsilon _{R}|0\rangle
_{2_{L}}|1\rangle _{2_{R}} \text{ .}  \label{phiLR}
\end{equation}%
where $|0\rangle _{2_L,2_R}, |1\rangle _{2_L,2_R}$ refers to the
Raman fields $2_L,2_R$ with 0,1 photons, respectively. Here, $\eta _{2}=\Delta
\beta _{r}+\Delta \gamma _{2}$, where $\Delta \beta
_{r}$ is the phase difference of the read beams at the \textit{L},\textit{\ R%
} ensembles, and $\Delta \gamma _{2}$ is the phase difference
acquired by the $2_{L},2_{R}$\ fields in propagation from the
ensembles to the beam splitter $BS_{2}$ in Fig. 1b. In our
experiment, the phases $\eta
_{1},\eta _{2}$ can be independently controlled and are actively stabilized by utilizing auxiliary fields at $%
1.06$ $\mu $m that copropagate along the paths of the write and read
beams and of the $1_{L},1_{R}$\ and $2_{L},2_{R}$ fields.

Of course, the states in Eqs.~\ref{LRent}-\ref{phiLR} are
idealizations that must be generalized to describe our actual
experiment \cite{duan01,vanenk05,duan02}. Specifically, the presence
of various sources of noise necessarily transforms these pure states
into mixed states. Eqs. \ref{LRent}, \ref{phiLR} also neglect the
vacuum contribution as well as higher-order terms, which are
intrinsic to DLCZ protocol and which otherwise arise due to diverse
experimental imperfections. Moreover, the above analysis assumes
that all excitations are in the correct \textquotedblleft
modes\textquotedblright (both for optical fields and for
the collective atomic \textquotedblleft spin flips\textquotedblright), that excitations of the  ensembles map $1$%
-to-$1$ to photons in fields 1 and 2, and
that diverse sources of background light are absent.

The procedure that we have devised to provide a robust, model independent determination of entanglement is based upon quantum tomography of the $2_{L},2_{R}$ fields (see Appendix B for details).
Since entanglement cannot be increased by local operations on either
of the two ensembles, the entanglement for the state of the
ensembles will be always greater than or equal to that
measured for the state of the light fields.
Specifically,
conditioned upon a detection at $D_{1a}$ or $%
D_{1b}$, we consider the density matrix%
\begin{equation}
\tilde{\rho}_{2_{L},2_{R}}=\frac{1}{\tilde{P}}\left(
\begin{array}{cccc}
p_{00} & 0 & 0 & 0 \\
0 & p_{01} & d & 0 \\
0 & d^{\ast } & p_{10} & 0 \\
0 & 0 & 0 & p_{11}%
\end{array}%
\right) ,  \label{rhotilde}
\end{equation}%
which is written in the basis $|n\rangle _{2_{L}}|m\rangle _{2_{R}}$%
, with the number of photons $\{n,m\}=\{0,1\}$. $p_{ij}$ is then the probability to find $i$ photons in mode $2L$ and $j$ photons in mode $2R$, and $d$ gives the coherence between the $|1\rangle _{2_{L}}|0\rangle
_{2_{R}}$ and $|0\rangle _{2_{L}}|1\rangle
_{2_{R}}$ states. $\tilde{\rho}%
_{2_{L},2_{R}}$ is obtained from the full density matrix $\rho
_{2_{L},2_{R}} $ by restricting it to the subspace where there is
at most one photon in each mode, with then
$\tilde{P}=p_{00}+p_{01}+p_{10}+p_{11}$.
Since the concurrence $C(\tilde{\rho}_{2_{L},2_{R}})$ for $\tilde{\rho}%
_{2_{L},2_{R}}$ provides a lower bound for the concurrence $C(\rho
_{2_{L},2_{R}})$ for $\rho _{2_{L},2_{R}}$ [$C(\rho
_{2_{L},2_{R}})\geq \tilde{P}C(\tilde{\rho}_{2_{L},2_{R}})$], we
devise measurements to deduce the various components of
$\tilde{\rho}_{2_{L},2_{R}}$. The concurrence $C(\tilde{\rho}_{2_{L},2_{R}})$ can then be
calculated from Eq. \ref{rhotilde} by way of Ref.
\cite{wooters},
\begin{equation} 
\tilde{P} \, C=\max(2|d|-2\sqrt{(p_{00}p_{11})},0)\text{ .}  \label{C}
\end{equation} 
The entanglement of formation $E$ follows directly from $C$, where
$E$ and $C$ both range from 0 to 1 for our system and $E$ is a
monotonically increasing function of $C$ \cite{wooters}.

\begin{table}[th]
\begin{center}
\begin{tabular*}{4 in}{|c@{\extracolsep{\fill}}cc|}
\hline \textbf{Probability} & \textbf{$D_{1a}$} & \textbf{$D_{1b}$}
 \\ \hline
$p_{00}$ & $0.98510\pm 0.00007$ & $0.98501\pm 0.00007$  \\
$p_{10}$ & $(7.38\pm 0.05)\times 10^{-3}$ & $(6.19\pm 0.04)\times 10^{-3}$ \\
$p_{01}$ & $(7.51\pm 0.05)\times 10^{-3}$ & $(8.78\pm 0.05)\times10^{-3}$ \\
$p_{11}$ & $(1.7\pm 0.2)\times 10^{-5}$ & $(1.9\pm 0.2)\times
10^{-5}$ \\ \hline
\end{tabular*}%
\end{center}
\caption{Diagonal elements of the density matrix $\tilde{\rho}%
_{2_{L},2_{R}}$ [Eq. (3)], deduced from the records of
photo-electric counts. The values of $p_{ij}$ are referenced to the
location of detectors $D_{2a,2b,2c}$, and were obtained by
considering unit detection efficiency, which gives a more
conservative (i.e., smaller) lower bound for the concurrence than
the actual (larger) field concurrence for finite efficiency $<1$.
See Appendix A for further details.}
\end{table}

As a first step in the determination of $C$ we measure the
diagonal elements $p_{ij}$. As shown in Fig.~1b, the field-2
output of each ensemble is directed to different sets of detectors
in order to record photon counting probabilities for the fields
$2_{L},2_{R}$ separately. From the record
of photoelectric counting events, we then deduce the diagonal elements of $%
\tilde{\rho}_{2_{L},2_{R}}$, which are listed in Table 1. From
Eq.~\ref{C} and noting that $|d|^2 \leq p_{10}p_{01}$, a necessary
requirement for $C>0$ is that there be a suppression of two-photon
events relative to the square of the probability for single-photon
events for the fields $2_{L},2_{R}$  i.e., $h_{c}^{(2)}\equiv
p_{11}/(p_{10}p_{01})<1$. For our measurements, we find
$h_{c}^{(2)}=0.30 \pm 0.04$ for events conditioned on detection at
$D_{1a}$, and $h_{c}^{(2)}=0.35 \pm 0.04$ for events conditioned on
$D_{1b}$~\cite{chou04}. In contrast, for non-conditioned events, we
find $h_{nc}^{(2)}=0.99 \pm 0.04$.

The second step in our tomography protocol is to determine the
coherence term $d$ in Eq. \ref{rhotilde}, which we accomplish by
adding a relative phase shift $\varphi $\ for the fields
$2_{L},2_{R}$, and by combining them at the beam splitter $BS_{2}$
shown in Fig. 1b. By recording the conditional count rate after the
beam splitter as function of $\varphi $, one can measure an
interference fringe with a visibility $V$, with then $|d|$ following
from $V$ and the $p_{ij}$. Roughly, for 50/50 beam splitters and
neglecting higher order terms (that are employed in our actual analysis), we would have $|d| \cong V
(p_{10}+p_{01})/2$.

\begin{figure}[th]
\begin{center}
\includegraphics[width=12cm]{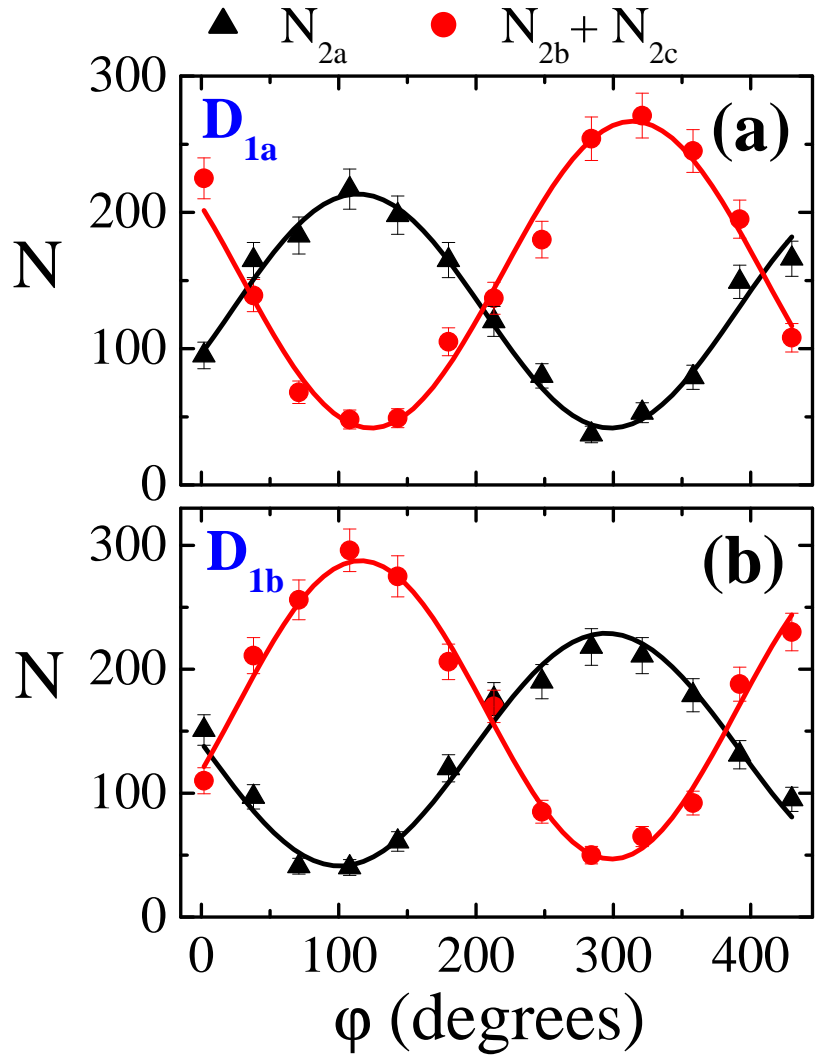}
\end{center}
\vspace{-0.7cm} \caption{Coherence between the atomic ensembles
$L,R$ is induced by a
measurement event of the fields $1_{L},1_{R}$ at detector $D_{1a}$ or $%
D_{1b} $. Shown is the number of coincidences $N_{2a}$ (circles) and $%
N_{2b}+N_{2c}$ (triangles) recorded by the respective detectors
$D_{2a,2b,2c}$ for the fields $2_{L},2_{R}$ with the interferometer
arrangement of Fig.
1b as a function of the relative phase $\protect\varphi $. In (a) $%
N_{2a,2b,2c}$ are conditioned upon a detection event at $D_{1a}$
with no count at $D_{1b}$, while in (b) $N_{2a,2b,2c}$ are
conditioned upon an event
at $D_{1b}$ with no count at $D_{1a}$. At each setting of $\protect\varphi $%
, data are acquired for $150$ s with a detection window of width
$190$ ns. Although the interference fringes have comparable
visibility, the different
sizes arise from unequal quantum efficiencies for detectors $D_{2a}$ and $%
D_{2b,2c}$ (see Appendix A). The visibility values
are obtained from an average of the visibilities of the red and
black curves, respectively. Error bars reflect $\pm $\ one standard
deviation due to the finite number of counts.} \label{fig2}
\end{figure}

Figure 2 shows conditional counts $N_{2a},N_{2b}+N_{2c}$ as
functions of  $\varphi $. These data demonstrate that the
indistinguishable character of measurement events at detectors
$D_{1a}$ (Fig.2a) and $D_{1b}$ (Fig.2b) induces a high degree of
coherence between the otherwise independent ensembles $L,R$
\cite{duan01,matsukevich04}. Indeed, we deduce visibilities
$V_{1a}=(70\pm 2)\%$ and $V_{1b}=(71\pm 2)\%$ for the associated
conditional states.

A notable feature of these results is that the interference fringes
have relative phase $\pi $ for the cases of detection at
$D_{1a},D_{1b}$, in agreement with Eqs. \ref{LRent}, \ref{phiLR}. We
observe similar fringes if the phase $\eta_1$ between the write
beams is varied instead of $\varphi $. Moreover, if the fields
$1_{L},1_{R}$\ are combined at the beamsplitter $BS_{1}$ with orthogonal polarizations (by way of the half-wave plate in Fig. 1a), we find that the visibility from
interference of fields $2_{L},2_{R}$ drops to near
zero, since in this case, there is no longer measurement-induced
entanglement associated with quantum interference for detection of fields $1_{L},1_{R}$ (see Appendix A).

With Eq. \ref{C}, the measured values for the visibility $V$ and  for the various $p_{ij}$ are sufficient to deduce a lower bound for the concurrence
$C$\ for
the field state $\tilde{\rho}_{2_{L},2_{R}}$ at the location of detectors $%
D_{2a,2b,2c}$, with {\it no correction for detection
efficiencies or propagation losses}, we find
\begin{equation}
C_{1a}(\tilde{\rho}_{2_{L},2_{R}})=(2.4\pm 0.6)\times 10^{-3}>0\;,\;C_{1b}(%
\tilde{\rho}_{2_{L},2_{R}})=(1.9\pm 0.6)\times 10^{-3}>0\,,
\label{C1ab}
\end{equation} conditioned upon detection at either $D_{1a}$ or $D_{1b}$. This
conclusively demonstrates a nonzero
degree of entanglement between the ensembles, albeit with the concurrence $%
C_{L,R}$ small. The small difference between the concurrence for the
states conditioned on $D_{1a}$ or $D_{1b}$ can be explained by an
asymmetry in $BS_w$ (see Appendix A).

\begin{figure}[t]
\begin{center}
\includegraphics[width=10cm]{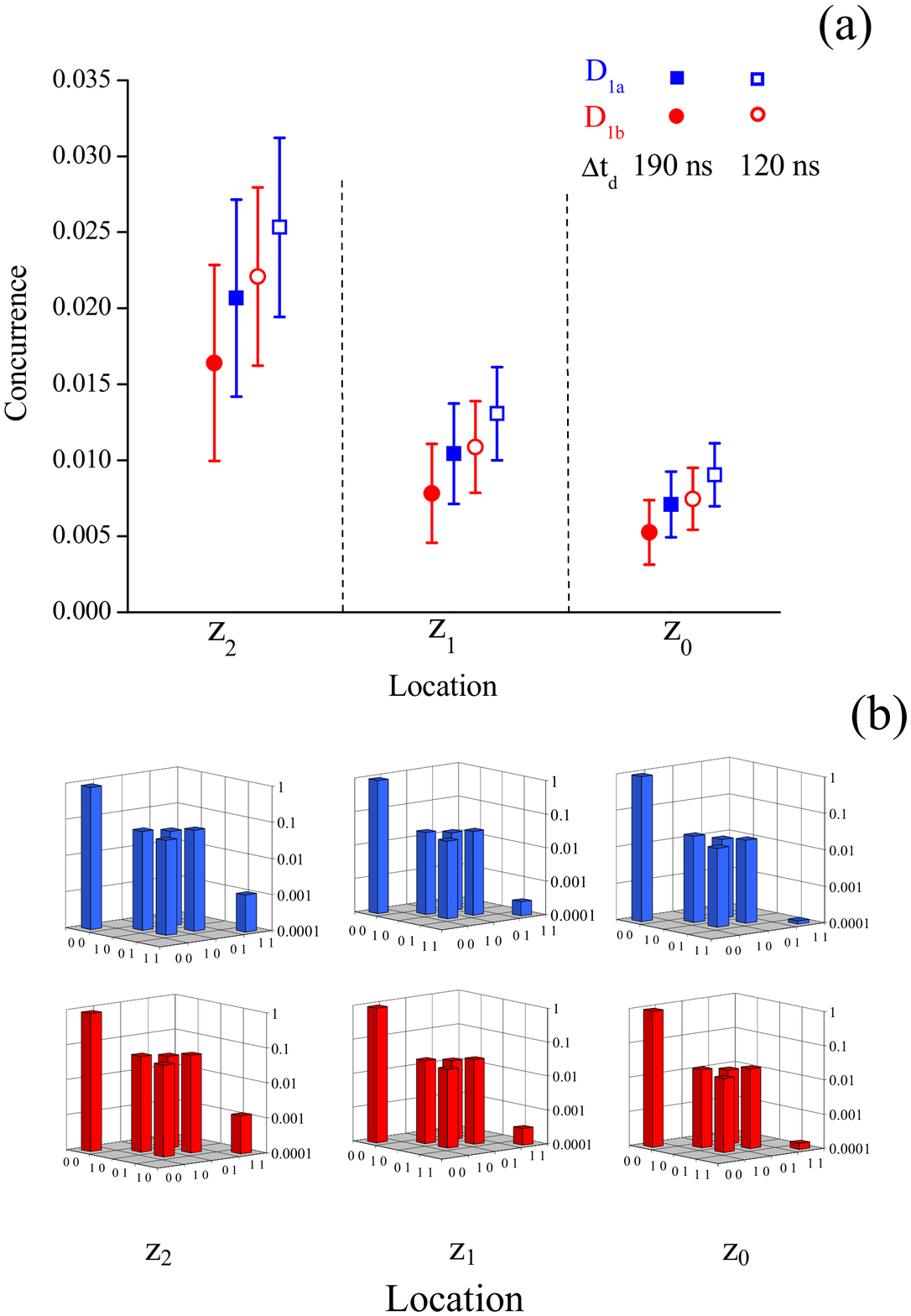}
\end{center}
\caption{The results of our measurements for the concurrence
$C^{z_{i}}$ (a) and density matrix
$\tilde{\protect\rho}_{2_{L},2_{R}}^{z_{i}}$ (b) are shown at the
three locations $z_{i}$ indicated in Fig. 1b. At each location, two
pairs of results are given corresponding to the measurement-induced
state created from detection at $D_{1a}$ and at $D_{1b}$, and taking
into account the efficiency of the detectors and propagation losses.
(a) Concurrence $C$, for two different detection windows $\Delta
t_d$ at $D_{2a,2b,2c}$. Filled symbols are for  $\Delta t_{d}=190$
ns, enough to acquire the whole temporal wavepacket of field 2. Open
symbols are for $\Delta t_{d}=120 $ ns. We see then that the degree
of entanglement can be further enhanced, similar to the increase of
nonclassical correlations between fields 1 and 2 reported in
Ref.~\cite{polyakov04} for specific detection windows for these
fields. All values shown in this figure, including the ones for
$z_{0}$, are already corrected for
the efficiencies of the detectors. Error bars reflect $\pm$ 1 standard deviation, taking into account the finite number of counts and the uncertainties in the efficiency and propagation loss. (b) Density matrix $\tilde{\protect\rho}%
_{2_{L},2_{R}}^{z_{i}}$ given in the basis $|n\rangle
_{2_{L}}|m\rangle _{2_{R}}$ corresponding to Eq.
\protect\ref{rhotilde}, with $\{n,m\}=\{0,1\}$ and $\Delta t_d = 190$~ns.} \label{fig3}
\end{figure}

Beyond the firm lower bound given by Eq. \ref{C1ab}, we can make a
better estimate of the degree of entanglement $C_{L,R}$\ between the
$L,R$ ensembles by way of detailed measurements of the propagation
efficiencies from the atomic ensembles to the plane $z_{0}$\ of the
detectors shown in
Fig. 1b (see Appendix A). Figure 3 gives an inference of the density matrix $%
\tilde{\rho}_{2_{L},2_{R}}^{z_{i}}$ and thereby of the concurrence $%
C^{z_{i}}(\tilde{\rho}_{2_{L},2_{R}}^{z_{i}})$ at $z_{0}$ and at two
other locations $z_{i=1,2}$\ along the path from the ensembles to
the detectors (see Fig. 1b), assuming a constant visibility. Generally, $C$ increases in
direct correspondence to the reduced level of losses for the
$2_{L},2_{R}$ fields at locations closer to the ensembles. At
location $z_{2}$ corresponding to the output edges of the atomic
ensembles, we find the result
\begin{equation}
C_{L,R}^{1a}\geq
C_{1a}^{z_{2}}(\tilde{\rho}_{2_{L},2_{R}}^{z_{2}})\simeq
0.021\pm 0.006>0\;,\;C_{L,R}^{1b}\geq C_{1b}^{z_{2}}(\tilde{\rho}%
_{2_{L},2_{R}}^{z_{2}})\simeq 0.016\pm 0.006>0\,.  \label{ELR}
\end{equation}
To move beyond this result, we need more detailed information about
the efficiencies $\xi _{L,R}$ with which stored excitation in the
atomic ensembles is converted to the propagating light fields
$2_{L},2_{R}$. Our earlier measurements included comparisons to a
simple model \cite{chou04}\ and allowed an inference $\xi _{L,R}\sim
0.10\pm 0.05$. The measurement of the losses together with the
values of $p_{ij}$ at the detectors yields $p_{10}+p_{01}\approx
11\%$ at the output of the ensembles ($z_{2}$
plane) for our current experiment. This value together with the estimated $%
\xi _{L,R}$ then indicates that $p_{00}\rightarrow 0$ for the
conditional state $\rho _{L,R}$ of the ensembles, so that
$C_{L,R}\approx V\approx 0.7$, suggesting that $\rho _{L,R}$ is
close to the ideal entangled state of Eq. \ref{LRent}. The low
measured values for the entanglement between fields $2_{L},2_{R}$
apparently are principally a consequence of the low readout
efficiency $\xi _{L,R}$ of the atomic excitation. We stress that
this inference of $C$ for the state inside the ensembles must be
confirmed by subsequent experiments and is offered here to provide
some insight into future prospects for quantum protocols with
entangled ensembles. This also emphasizes that a central point in
subsequent work should be the improvement of $\xi _{L,R}$.

In conclusion, we have achieved entanglement between a pair of
atomic ensembles separated by $2.8$ m, with the entangled state
involving one spin excitation within a collective system of roughly
$10^{5}$ atoms at each site $L,R$. The entangled state is generated
by and conditioned upon an initial detection event, and is thus
produced in a probabilistic fashion. However, this initial event
heralds unambiguously the creation of an entangled state between $L,R$
ensembles, which is physically available for subsequent utilization,
as, for example, by mapping to propagating optical fields, which can be in
principle accomplished with high efficiency. We
emphasize that our measurements relate to an actual physical state of the $%
L,R$\ ensembles and of the $2_{L},2_{R}$ fields, and are not an
inference of a state based upon post selection. Our work provides
the first example of a stored atomic entangled state that can be
transfered to entangled light fields, and significantly extends
laboratory capabilities for entanglement generation, with now
entangled states of matter stored with separation a hundred-fold
larger than was heretofore possible for continuous quantum variables
and a thousand-fold larger than for qubits. With our current setup,
we have demonstrated $\Delta t_{s}\simeq 1$ $\mu $s for storing
entanglement. However, this should be readily extended to $\Delta
t_{s}\simeq 10$ $\mu $s, and new trapping schemes have the potential
to lead to $\Delta t_{s}\simeq 1$ s \cite{felinto05}. The distance
scale for separating the $L,R$\ ensembles is limited by the length
$l_{0}\simeq 2$ km for fiber optic attenuation at our write
wavelength $852$ nm. Extensions to scalable quantum networks over
larger distances requires the realization of a quantum repeater
\cite{duan01}, for which we have now laid the essential
foundation.\bigskip

\bigskip

\textbf{Acknowledgement -} This research is supported by the
Advanced Research and Development Activity (ARDA), by the National
Science Foundation, and by the Caltech MURI Center for Quantum
Networks. D.F. acknowledges financial support by CNPq (Brazilian
agency). H.d.R. acknowledges financial support by the Swiss National
Science Foundation. SJvE thanks Lorenz Huelsbergen for his
assistance in computer matters.

\bigskip

\appendix

\section{Experimental Details}

\subsection{Atomic ensembles and optical pulses}

Each of the $L,R$
atomic ensembles is obtained from Cesium atoms in a magneto-optical
trap (MOT) \cite{chou04,felinto05}. Measurements are carried out in
a cyclic fashion consisting first of a period of cooling and
trapping to form the MOT, followed by an interval during which the
magnetic fields for the MOT\ are switched off. After waiting $3$ ms
for the magnetic field to decay~\cite{felinto05}, we initiate a
sequence of measurement trials, where for each trial the atoms are
initially prepared in level $|g\rangle$. The write pulse is at 852
nm, with a duration of $150$ ns and is detuned 10 MHz below the
$|g\rangle \rightarrow |e\rangle $ transition. The read pulse is at
894 nm, with a duration of 130 ns and is resonant with the
$|s\rangle \rightarrow |e^{\prime }\rangle$ transition. At the end
of each trial, the sample is pumped back to level $|g\rangle $ by
illuminating the atomic cloud with trapping and repumping light for
$0.7 \mu $s and 1 $\mu $s respectively, and then a new trial is
initiated with period of $3$ $\mu $s. The total duration for a
sequence of measurement trials is $5$ ms, after which the
measurement interval is terminated and a new MOT\ is formed in
preparation for the next sequence of trials.

\subsection{Losses and efficiencies}

In order to infer the concurrence of the fields at different
locations in our experimental apparatus it is important to
characterize the losses in the different components of the
communication channel of field 2. Our measurements for such losses
are given in the table below for the pathways starting from each of
the atomic ensembles. Immediately after the ensembles, we separate
the classical pulses from fields 1 and 2 in the single-photon
level. This procedure is explained in detail in Refs. \cite{kuzmich03,chou04}%
. The losses coming from this stage are due mainly to passage
through paraffin coated vapor cells, which have transmission $\alpha
_{fc}$. After the filtering process, field 2 is coupled to a
polarization maintaining fiber that carries it to the detection
region. The coupling efficiency is denoted by $\alpha _{c}$. In the
detection region, it is important first to filter the 1.06~$\mu $m
light that propagates together with field 2, and which is used to
actively lock the read interferometer when $BS_{2}$ is inserted. In
order to filter the 1.06~$\mu $m light, field 2 comes out of the
fiber and pass trough a bulk low-pass filter that reflects 1.06~$\mu
$m light and transmits with high efficiency at 894~nm. At this
stage, we also include an extra filter for 852~nm, to cut any
residual light from the write process, and then field 2 is coupled
again to a fiber, this time a multi-mode one. This whole process of
passing through the band-pass filters and recoupling to fiber is
called generally \textquotedblleft filter for 1064~nm", and is
characterized by a transmission $\alpha _{f}$. Finally, the detector
efficiencies are given by $\alpha _{APD}$. Note that a single number
is given for the pair of detectors $D_{2b,2c}$, since they both have
measured efficiencies of 40\%.

\begin{table}[th]
\begin{center}
\begin{tabular*}{5.4 in}{|c@{\extracolsep{\fill}}cccc|}
\hline \textbf{Description} & \textbf{Symbol} & \textbf{Value for
ensemble L} & \textbf{Value for ensemble R} & \textbf{Error} \\
\hline\hline
filter cell & $\alpha _{fc}$ & 0.80 & 0.80 & 0.02 \\
fiber coupling & $\alpha _{c}$ & 0.70 & 0.65 & 0.02 \\
1064 nm filter & $\alpha _{f}$ & 0.70 & 0.70 & 0.02 \\
detector & $\alpha _{APD}$ & 0.32 & 0.40 & 0.02 \\
Total & $\alpha $ & 0.13 & 0.15 &  \\ \hline
\end{tabular*}%
\end{center}
\par
{TABLE 1: List of efficiencies associated with photon 2 propagation
and detection.}
\end{table}

\subsection{Suppression of interference between the $2_{L},2_{R}$ fields for
distinguishable \newline detection events from the fields
$1_{L},1_{R}$}

\begin{figure}[th]
\begin{center}
\includegraphics[width=12cm]{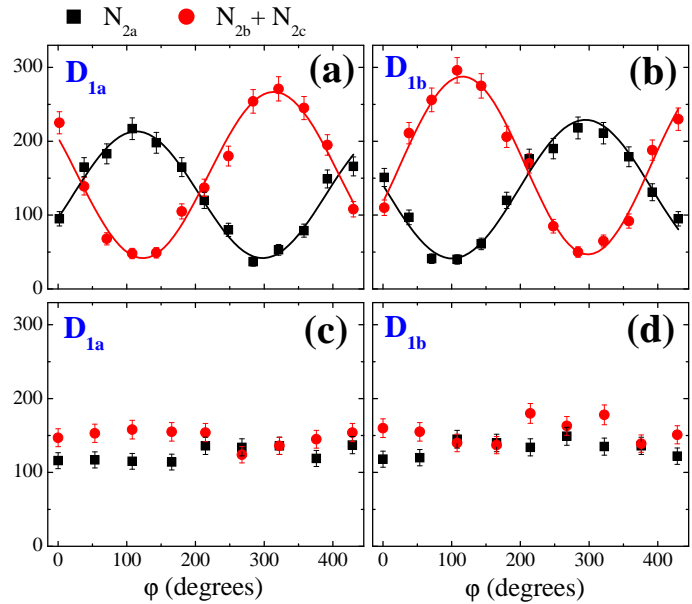}\\[0pt]
\end{center}
\par
\vspace{-4cm} \caption{Number of coincidences $N_{2a}$ (squares) and
$N_{2b}+N_{2c}$ (circles) recorded by the respective detectors
$D_{2a,2b,2c}$ for the fields
$2_{L},2_{R}$ with the interferometer arrangement of Fig.~1(b) as a function of the relative phase $\protect\varphi $%
. Frames (a) and (b) are the same as frames (a) and (b) of Fig.~2, i.e., they show the interference fringe
between fields $2_{L},2_{R}$ as a result of combining fields
$1_{L},1_{R}$ in an approximately indistinguishable fashion with
parallel polarizations. Frames
(c) and (d) show the results of the same measurement on fields $2_{L},2_{R}$%
, but now with fields $1_{L},1_{R}$ combined with orthogonal
polarizations. At each setting of $\protect\varphi $, data are
acquired for $150$ s with a detection window of width $190$ ns.}
\label{fringe}
\end{figure}

In our experiment, entanglement is generated by quantum interference
between the fields 1 emitted by the ensembles, that are combined at
a beam-splitter. For this interference to occur, the two fields must
be indistinguishable, such that no information can be obtained about
the origin of the photons. A good way to illustrate the importance
of this overlap is to render the photons distinguishable, for
instance by combining the two fields 1 with orthogonal
polarizations. In this way, information about the origin of the
detected photon is near maximal, and the degree of
measurement-induced entanglement should be significantly reduced (to
zero in the ideal case).
The results of such a measurement are shown in Fig. \ref{fringe}. Fig.~\ref%
{fringe}(a,b) shows the interference fringes obtained when the
photons of
field 1 are combined with parallel polarizations (same fringes as in Fig.~\ref{fig2}), while in Fig.~\ref{fringe}(c,d), the photons from field 1
are combined with orthogonal polarizations. In the latter case, the
visibility drops to near zero, and there is no entanglement between
the two ensembles. The residual oscillation in the conditional count
rate can be explained by the finite polarization extinction ratio in
our polarization maintaining fibers. The fibers used in our
experiment have a measured extinction ratio of 28~dB between their
two orthogonal propagation modes. This can lead to a residual
visibility of 8 $\%$, which is compatible with the amplitude of the
residual oscillation in Fig.~\ref{fringe}(c,d).

\subsection{Asymmetry in the creation of the states conditioned on $D_{1a}$
and $D_{1b}$}

The difference in the two sets of probabilities $%
(p_{01}^{(1a)},p_{10}^{(1a)})$ and $(p_{01}^{(1b)},p_{10}^{(1b)})$
results from an asymmetry in the beam splitter $BS_{1}$ for
detection of the write fields $1_{L},1_{R}$, with a measured ratio
of transmission to reflection $T/R=0.85$. Hence, in addition to the
$\pm $ sign in Eq.~1 of set by detection at $D_{1a}$
or $D_{1b}$, the relative amplitudes for the conditional state can
also differ, resulting in different values for the concurrence.  We
expect the ratio
$(p_{01}^{(1a)}/p_{10}^{(1a)})(p_{01}^{(1b)}/p_{10}^{(1b)})^{-1}$ to
be $(T/R)^2=0.73$, which agrees well with the measured value
$(7.51/7.38)(8.78/6.19)^{-1}=0.72$.

\vspace{2.4cm}

\subsection{Verification of single excitation regime}
\label{high}

Apart from the parameter $h_c^{(2)}$ discussed previously, another measure of the
single-photon character of field 2 is given by the evaluation of the
function $\tilde{g}_{12} = \tilde{p}_{12}/(\tilde{p}_1\tilde{p}_2)$
for each ensemble separately~\cite{kuzmich03,chou04}, where
$\tilde{p}_{12}$ is the joint probability of detecting a photon in
field 1 and another in field 2, and $\tilde{p}_i$ is the probability
of detecting a single photon in field $i$. For the situation of our
measurement, we obtained $\tilde{g}_{12}$ values of about 9 and 11
for ensembles $R$ and $L$, respectively, which is then an indication
of the single-photon character of field 2 emitted by both ensembles
separately, as discussed in~\cite{chou04}.

\subsection{Phase stabilization.} 

The $\approx\,$12~m path lengths for the two interferometers
formed by $(BS_{w},BS_{1})$ and by $(BS_{R},BS_{2})$ shown in
Figures 1a,b, respectively, are held constant by
injecting an iodine stabilized Nd:YAG laser into the fiber beam
splitters $BS_{w},BS_{R}$ for the write and
read beams. The fields at $1.064$ $\mu $m emerging from beam splitters $%
BS_{1},BS_{2}$ are directed to separate sets of detectors (not shown
in the figure), whose outputs are utilized to stabilize the relative
path lengths of the respective Mach-Zehnder interferometers by
feedback to piezoelectric transducers on which are mounted mirrors
in the paths of the write and read beams in the $L$ arms of the
respective interferometers. Since the write and read beams are
combined together before being focused into the ensembles, the
interferometers share a common free space path when they cross the
atoms. In order to control them independently and minimize
crosstalk, the two interferometers are addressed alternately at a
rate of 400~kHz.

\subsection{Normalization and detector configuration}

The role of the three detectors in our experiment is as a
second check on the order of magnitude of the two-photon events. Our
main checks were discussed in the manuscript and in section~\ref{high} of this
Appendix. It is interesting to obtain
directly $p_{02}$ and have an idea of its effect on
$p_{01}$. For our measurement, we obtained, with $D_ {2b}$ and $D_{2c}$,
$p_{02}=(2.2 \pm 0.4)\times10^{-5}$ conditioned on detector
$D_{1a}$, and $p_{02}=(2.4 \pm 0.4)\times10^{-5}$ conditioned on
detector $D_{1b}$. Note that these values are so small that if taken into
account, the correction for
$p_{10}$ and $p_{01}$ would be negligible. In order to
simplify our analysis and inversion algorithm, we
consider the three detectors as an effective set of two detectors.
This is done by simply
adding the coincidence events between $D_{2b}$ and $D_{2c}$ to
the sum $N_{2b}+N_{2c}$, in the same way that would result from the use of a
non-number-resolving detector.
In this case, the measured normalization constant $\tilde{P}$ is equal to
one, since all measured events contribute to one
of the elements of the restricted density matrix. If we had used more
detectors,
the discrepancy of $\tilde{P}$ from unity would be on the order of $p_{02}$
and negligible compared to the experiment accuracy.

\section{Theory}

\subsection{Entanglement}

For convenience of description we assume the two atomic ensembles
$L$ and $R$ to be in the hands of Alice and Bob, respectively. The
state of the two
ensembles conditioned on a click of one of the two detectors $D_{1a}$ or $%
D_{1b}$ (see Figure~1b of main text) is mapped onto a state of
multiple field modes belonging to Alice and multiple modes belonging
to Bob. Because the mapping involves only \emph{local} operations by
Alice and Bob, the entanglement (in particular, the entanglement of
formation) between their systems cannot increase on average
\cite{localop}. Hence the entanglement found between Alice's and
Bob's field modes is a lower bound on the entanglement between the
atomic ensembles. We will use this type of reasoning several times
here: certain experimental procedures can be exactly mimicked by
imagining Alice and Bob performing LOCC (local operations and
classical communication): those operations can only decrease the
entanglement we find. We also sometimes (lower) bound the
entanglement analytically using quantities that are more
straightforward to measure in the laboratory. That way, we can
unambiguously determine the presence of entanglement between the two
ensembles, even if we might underestimate its actual magnitude.

On each side there is one main mode (a traveling continuous-wave
mode) into which photons are emitted predominantly \cite{duan01}.
Those modes we denote by $2_{L}$ and $2_{R}$. Other modes may be
populated with very small probability, but in the analysis we assume
all detector clicks arise from modes $2_{L}$ and $2_{R}$. In the
experiment this reduction from multiple to single modes is mainly
accomplished by the use of single mode fibers, which filter out
different spatial modes. This is a procedure that can be exactly
mimicked by Alice and Bob performing that same spatial filtering on
their local modes and hence can only decrease the actual
entanglement.

We also assume that never more than 2 photons populate each mode.
This is an excellent approximation on its own (and is supported by
our measurements), but more importantly, this assumption corresponds
to lower bounding the entanglement, as detailed below.

We furthermore assume that all off-diagonal elements of the density
matrix between states with different numbers of photons vanish. This
is a valid assumption when one takes into account that that phase
can only be defined relative to a reference laser beam shared by
Alice and Bob. Tracing out that laser field sets the off-diagonal
elements to zero. Indeed, the experiment makes no use of knowledge
of the phases of the various lasers used. Moreover, this can only
underestimate the entanglement, since tracing out the laser modes
can be exactly mimicked by Alice and Bob performing local operations
that makes all those off-diagonal elements zero. Namely, they could
each apply a random phase shift to their modes, such that the phase
difference is fixed (this requires classical but not quantum
communication), and subsequently ignore the information about the
individual phase shifts. The phase difference is equal to the phase
$\varphi $ introduced in the main text.

This then leaves us with a density matrix of the form
\[
\rho _{2_{L},2_{R}}=\left(
\begin{array}{cccccc}
p_{00} & 0 & 0 & 0 & 0 & 0 \\
0 & p_{01} & d & 0 & 0 & 0 \\
0 & d^{\ast} & p_{10} & 0 & 0 & 0 \\
0 & 0 & 0 & p_{11} & e & f \\
0 & 0 & 0 & e^{\ast} & p_{02} & g \\
0 & 0 & 0 & f^{\ast} & g^{\ast} & p_{20} \\
&  &  &  &  &
\end{array}%
\right) .
\]%
We can bound the entanglement of formation of this state by
\begin{equation}
E(\rho _{A,B})\geq \tilde{P}E(\tilde{\rho}_{2_{L},2_{R}})
\end{equation}%
where
\begin{eqnarray}
\tilde{P} &=&p_{00}+p_{01}+p_{10}+p_{11} \\
\tilde{\rho}_{2_{L},2_{R}} &=&\frac{1}{\tilde{P}}\left(
\begin{array}{cccc}
p_{00} & 0 & 0 & 0 \\
0 & p_{01} & d & 0 \\
0 & d^{\ast } & p_{10} & 0 \\
0 & 0 & 0 & p_{11} \\
&  &  &
\end{array}%
\right)
\end{eqnarray}%
One obtains this bound by considering the effects of two local
operations by Alice and Bob consisting of measuring whether each
mode has more than 1 photon or not and communicating this result one
to the other. We treat this step explicitly and separately from the
very similar step mentioned above [where cases with more than two
photons are filtered out], in order to remind ourselves we do have
to keep track of the total probability to find more than 1 photon in
one of the modes, $1-\tilde{P}$. Also, we note explicitly this step
does not correspond to any procedure in our experiment, but is just
an analytic tool to bound the entanglement and express it in terms
of quantities that can be easily determined without too large
uncertainty (unlike higher-order matrix elements such as $p_{12}$,
etc).

\subsubsection{Detection window}

In Fig.~3, we show results for a smaller detection
window that give a higher estimation for the amount of entanglement
between the fields. The use of a smaller detection window can also
be understood as a reduction from multiple to single modes by
filtering out different temporal modes. This is a procedure that can
be exactly mimicked by Alice and Bob performing that same temporal
filtering on their local modes and hence can only decrease the
actual entanglement (note though, that the \emph{estimate} of
entanglement can be increased by this procedure, even though the
actual entanglement decreases. The estimated value just gets much
closer to the actual value than when not filtering out extraneous
modes).

For the possibly remaining multiple modes in the smaller time
window, we may suppose Alice and Bob each apply the following
fictitious transformation
\[
|n_1\rangle|n_2\rangle\ldots\rightarrow |\sum_i n_i\rangle
|n_1,n_2,\ldots\rangle_M
\]
to their local modes. This transformation collects all photons in
one mode (the first ket), and keeps track of where they came from in
a separate system $M$ (for ``memory''), such that the transformation
is unitary. This transformation, therefore, leaves the entanglement
between Alice's and Bob's mode unchanged. Our detectors not being
able to distinguish different modes within the same time window then
boils down to tracing out the memory system, which can only decrease
the entanglement.

\subsection{Measurements}

All measurements are performed using Geiger-mode avalanche
photodiodes (APDs). We assume there are only two outcomes of a
photodetection measurement, corresponding to no click or some
nonzero number of clicks (indeed, that is the only information used
in the experiment). Thus, if the incoming mode $2_{L}$ is described
by an annihilation operator $a$ then the photodetector performs a
POVM of the form
\begin{eqnarray}
\Pi _{0} &=&\sum_{n\geq 0}(-1)^{n}\frac{a^{\dagger n}a^{n}}{n!}
\nonumber
\label{povm} \\
\Pi _{1} &=&I_{A}-\Pi _{0},
\end{eqnarray}%
with $I_{A}$ the identity on $2_{L}$. The corresponding
probabilities are denoted by $p_{0}$ and $p_{1}$,
\begin{equation}
p_{k}=\mathrm{Tr}\rho _{2_{L}}\Pi _{k},  \label{probs}
\end{equation}%
if $\rho _{2_{L}}$ is the state of mode $2_{L}$. To deduce
higher-order probabilities with $k\geq 2$, beam splitters can be
employed to divide and direct the mode $a$ to multiple detectors.

Joint probabilities for measurements on the two modes $2_L$ and
$2_R$ can be determined in a similar way if we introduce the
annihilation operator $b$ for mode $2_R$ and the corresponding
operators $\Pi_n^{A,B}$ for $n=0,1$ for the two modes. The operators
$\Pi_0$ and $\Pi_1$ above were written in normal order. Similarly,
joint measurement probabilities can be written as
\begin{equation}  \label{normal}
P_{mn}=\mathrm{Tr} \tilde{\rho}_{2_L,2_R} :\Pi^A_m\otimes\Pi^B_n:,
\end{equation}
where all annihilation operators $a$ and $b$ are written to the
right of all creation operators $a^{\dagger}$ and $b^{\dagger}$. For
example, this allows
one to include easily the effects of finite efficiencies of detectors: If $%
\eta$ is the efficiency of a detector and $a$ the annihilation
operator for
the mode impinging on the detector, we may replace $a\rightarrow \sqrt{\eta}%
a+\sqrt{1-\eta}v$ where $v$ acts on an auxiliary mode that is
assumed to be in the vacuum state. Terms with nonzero powers of $v$
then do not contribute to counting rates provided we evaluate these
by using a normally-ordered
expression. In that case, we can ignore $v$ and just replace $a\rightarrow%
\sqrt{\eta}a$. The same replacement can be used to take into account
losses during propagation.

The most straightforward way to determine $\tilde{\rho}_{2_L,2_R}$
consists of two stages. The first stage determines the diagonal
elements, the second the off-diagonal elements: From the measured
frequencies of joint detection events we obtain estimates for the
corresponding probabilities for those joint events in terms of the
underlying density matrix elements. Inverting these expressions
gives estimates on the elements of the density matrix.

\subsubsection{Diagonal elements}

Conditioned upon detection of an event at either detector $D_{1a}$
or $D_{1b} $, the diagonal elements of $\tilde{\rho}_{2_{L},2_{R}}$
were measured by the setup described in Figure~1(c). Photons in mode
$2_{L}$ are detected by a photodetector $D_{2a}$ but mode $2_{R}$ is
split on a 50/50 (approximately) beamsplitter and photons in the two
resulting modes are counted by photodetectors $D_{2b}$ and $D_{2c}$.
Starting with mode operators $a$ and $b$ for modes $2_{L}$ and
$2_{R}$, there are several transformations affecting $a$ and $b$.
Denoting by $a_{1}$ the mode operator
detected by detector $D_{2a}$, and by $b_{1}$ and $b_{2}$ those detected by $%
D_{2b}$ and $D_{2c}$, the transformations are simply
\begin{eqnarray}
a_{1} &=&\sqrt{\eta _{L}\eta _{1}}a  \nonumber  \label{modes} \\
b_{2} &=&\sqrt{\eta _{R}\eta _{2}}b/\sqrt{2}  \nonumber \\
b_{3} &=&\sqrt{\eta _{R}\eta _{3}}b/\sqrt{2}
\end{eqnarray}%
with $\eta _{L,R}$ indicating the total efficiency of propagating to
the
detectors, and $\eta _{1,2,3}$ the detector efficiencies of detectors $%
D_{2a},D_{2b},D_{2c}$. One then obtains expressions for the expected
joint count probabilities $p_{klm}$ with $k,l,m=0,1$ by substituting
(\ref{modes})
into (\ref{povm}) and (\ref{normal}). When acting on $\tilde{\rho}%
_{2_{L},2_{R}}$ the operators $a$ and $b$ are understood to be
$a\otimes I_{B}$ and $I_{A}\otimes b$.

In the end, we only consider the total number of counts in detectors
$D_{2a}$
and $D_{2b},D_{2c}$ together, leading to joint probabilities $Q_{mn}$ with $%
m=0,1$ and $n=0,1,2$,
\begin{equation}
Q_{mn}=\sum_{m}\sum_{s+t=n}P_{mst}.
\end{equation}%
Again, for these measurements as well as those in the next section,
the detection events at $D_{2a}$ and $D_{2b},D_{2c}$ are conditioned
upon an event at either $D_{1a}$ or $D_{1b}$. This gives expressions
for the expected detection probabilities $Q_{mn}$ as functions of
the diagonal elements $p_{mn}$ of the density matrix: conversely,
given the experimentally determined $Q_{mn}$ we invert the
expressions to obtain estimates for $p_{mn}$.

\subsubsection{Off-diagonal elements}

The off-diagonal elements are measured by including two extra
elements: one
is a phase shifter in one of the modes, say mode $2_{L}$, thus replacing $%
a\rightarrow \exp (i\varphi )a$. This phase is varied to produce the
interference pattern (\textquotedblleft fringes\textquotedblright )
of
Figure~2. Second is an extra 50/50 beamsplitter after the phase shifter [$%
BS_{2}$ in Fig. 1(d)]. One again easily arrives at simple
expressions for the operators $a_1$ and $b_1,b_2$ detected in terms
of $a$ and $b$, similar to that of (\ref{modes}). Just as before,
one then obtains expressions for the expected joint count
probabilities $p_{klm}$ with $k,l,m=0,1$ by
substituting these expressions for $a_1,b_1,b_2$  into (\ref{povm}) and (\ref%
{normal}). We thus find the joint detection probabilities $Q_{mn}^f$
for the
``fringes'' as a function of $d$ and the diagonal elements of $\tilde{\rho}%
_{2_L,2_R}$. Since we already obtained the diagonal elements in the
previous step, we then find an estimate of the diagonal element $d$
from the visibility of the fringes.

Note that we have also carried out measurements with detector
$D_{2a}$
replaced by a beam splitter and a pair of APDs, as for detectors $%
D_{2b},D_{2c}$. In this way, we confirm explicitly that higher order
events with $k\geq 2$ at $D_{2a}$ have a negligible impact on our
estimates of fringe visibility and of the probabilities
$p_{00},p_{10},p_{01},p_{11}$ that enter into the determination of
the concurrence $C$, in agreement with the independent assessment
from $D_{2b},D_{2c}$.

In addition to the above analysis we also performed a maximum
likelihood analysis of the density matrix. Given the actually
detected photon
statistics for both measurements of diagonal and off-diagonal elements \emph{%
together} one can, for each possible candidate density matrix $\rho
_{2_{L},2_{R}}$, calculate the probability that the actually
obtained measurement outcomes occur. Maximizing that probability
over all possible density matrices gives the most likely density
matrix. This estimate has been used as an additional check on the
inferred values quoted in the main text and their errors.

\end{document}